\documentclass[prb,twocolumn,showpacs,preprintnumbers,amsmath,amssymb]{revtex4}

\usepackage{graphicx}
\usepackage{bm}
\usepackage{amsmath}
\usepackage{amsfonts}
\usepackage{amssymb} 

\begin{document}

\title{One-dimensional Ising ferromagnet frustrated by long-range interactions
at finite temperatures}

\author{F. Cinti$^{1,2}$} 
\email{fabio.cinti@fi.infn.it}
\affiliation{
$^1$Dipartimento di Fisica, Universit\`a di Firenze I-50019 Sesto Fiorentino (FI), Italy \\
$^2$CNR-INFM S3 National Research Center, I-41100 Modena, Italy} 
\author{O. Portmann$^3$} 
\affiliation{$^3$Laboratorium f\"ur Festk\"orperphysik, Eidgen\"ossische
Technische Hochschule Z\"urich, CH-8093 Z\"urich, Switzerland}
\author{D. Pescia$^3$} 
\affiliation{$^3$Laboratorium f\"ur Festk\"orperphysik, Eidgen\"ossische
Technische Hochschule Z\"urich, CH-8093 Z\"urich, Switzerland}
\author{A. Vindigni$^3$}
\email{vindigni@phys.ethz.ch}
\affiliation{$^3$Laboratorium f\"ur Festk\"orperphysik, Eidgen\"ossische
Technische Hochschule Z\"urich, CH-8093 Z\"urich, Switzerland}

\date{\today}

\begin{abstract}
We consider a one-dimensional lattice of Ising-type variables where the ferromagnetic exchange interaction $J$ between neighboring sites is frustrated by a long-ranged anti-ferromagnetic interaction of strength $g$ between the sites $i$ and $j$, decaying as $\mid i-j\mid^{-\alpha}$, with $\alpha>1$. 
For $\alpha$ smaller than a certain threshold $\alpha_0$, which is larger than 2 
and depends on the ratio $J/g$, 
the ground state consists of an ordered sequence of  segments with equal length and alternating magnetization. The width of the segments depends on both $\alpha$ and the ratio $J/g$. Our Monte Carlo study shows that the on-site magnetization vanishes at finite temperatures and finds no indication of any phase transition.  
Yet, the modulation present in the ground state is recovered at finite temperatures in the two-point correlation function, which oscillates in space with a characteristic spatial period: The latter depends
on $\alpha$ and $J/g$ and  
decreases smoothly from the ground-state value as the temperature is increased.
Such an oscillation of the correlation function is exponentially damped over  
a characteristic spatial scale, the correlation length, which  asymptotically 
diverges roughly as the inverse of the temperature as $T=0$ is approached. 
This suggests that the long-range interaction causes the Ising chain to fall into a universality class consistent with an underlying continuous symmetry.
The $e^{\Delta/T}$ temperature dependence of the correlation length and the uniform ferromagnetic ground state, characteristic of the $g=0$ discrete Ising symmetry, are recovered for $\alpha >\alpha_0$.
\end{abstract}

\maketitle

\section{Introduction}
The competition between a short-ranged interaction favoring local order and a long-range interaction frustrating it on larger spatial scales is often used to explain 
pattern formation in chemistry, biology and physics~\cite{Seul,Muratov}. The role of the long-range interaction is to avoid  the global phase separation favored by the short-ranged interaction and promote a state of phase separation 
at mesoscopic or nano-scales. 
Thus, the long-range interaction is not, in general, a small perturbation~\cite{Lieb,Kiv1,Kiv2,Stariolo,Nussinov_1,Nussinov_2}, but must be considered as precisely as possible. From a computational point of view, this means that 
the frustrating interaction has to be accounted for by 
involving all the lattice sites in the computation, which in turn limits the actual system size that can be handled in e.g. Monte Carlo (MC) simulations~\cite{Debell,Viot_1,Viot_2,Viot_3,Cannas,Singer}. 
Few exact results on multi-scale, multi-interaction~\cite{Lieb,Kiv1} systems are present -- to our knowledge --
in literature. For one-dimensional systems rigorous proof of absence of a phase 
transition in the pure long-range antiferromagnetic model has been obtained~\cite{kerimov}. Besides, 
rigorous results concerning the  ground-state phase diagram can be found in Ref.~\onlinecite{Lieb}. 
Regarding two-dimensional lattice models with restricted spin orientation 
and dipole-dipole interaction competing with ferromagnetic nearest-neighbor
exchange interaction, Giuliani \textit{et al.}~\cite{Lieb2} showed that the ground state 
is periodic striped, while a zero-temperature reorientation transition (from in-plane to out-of-plane magnetization) occurs at 
a given relative strength of the short- and long-range interaction when are both antiferromagnetic. 
Finally, a generalization of this periodic ground state in some continuum versions has been rigorously
proved~\cite{cheng,Lieb3,muller}. 

In this paper, we perform MC simulations on a one-dimensional (1d) lattice with sites occupied by Ising-type classical variables assuming values $\sigma_j= \pm 1$. The nearest-neighbor sites interact by a short-ranged ferromagnetic interaction of strength $J$ which favors the same sign for two adjacent variables 
(in the language of magnetism the exchange interaction favors parallel alignment of neighboring spins). 
In addition, any two variables located at sites $i$ and $j$ interact by means of a long-range interaction of strength $g$ decaying according to a power law 
$\mid\!i\!-\!j\!\mid^{-\alpha}$ and favoring, instead, antiparallel alignment. In the present study, selected values of $\alpha>1$ and in the vicinity of 2 are investigated. This range turns out to be representative of the different physical regimes.  We are aware of the apparently academic nature of $i.$) a one-dimensional model and of $ii.$) this choice of values for $\alpha$. In fact, point charges interact via the Coulomb interaction, which has $\alpha =1$, while the dipolar interaction between two localized magnetic moments has $\alpha =3$. On the other side, imposing a mono-dimensional modulation to two- or three-dimensional arrangements of charges and spins (a symmetry often realized in experiments~\cite{Seul,Muratov,Ale,Oliver1}) produces an effective one-dimensional long-ranged interaction potential with an effective value of $\alpha $ which can differ from $1$ and $3$ respectively. As an example, elementary magnetic moments arranged into stripes and located on a two-dimensional array of sites interact with an effective, one-dimensional dipolar long-range interaction which, asymptotically, is proportional to $\mid\!i\!-\!j\mid^{-2}$~\cite{Ale}.
Accordingly, a systematic study for values of $\alpha $ in this range might reveal properties that can be used to explain physically relevant situations, such as those represented by the two-dimensional system of stripes quoted above
or similar models of frustration discussed in connection with electronic phase separation~\cite{Kiv2}. A 1d model has 
great computational advantages compared to its 2d and 3d counterparts, such as 
the possibility of simulating lattices of larger linear dimensions, 
which in turn allows larger modulation lengths than already reported~\cite{Debell,Cannas,Singer,Viot_1,Viot_2,Viot_3}, 
which are, indeed, closer to experimental situations. 
Later in the paper, we will single out the relevance of our results for understanding realistic spin and charged systems. 
Besides, variations of the 1d-Ising model including long-ranged potentials have been widely applied to biological problems~\cite{Bio_1}, such as protein folding~\cite{Bio_2} and helix-coil transitions~\cite{Alves_Hansmann_PRL}. 

This paper is organized as follows: In Section~\ref{gsp}, we introduce the model and its  known ground-state phase diagram~\cite{Lieb}  and then present our main results on the oscillatory character of the two-point correlation function, on the temperature dependence of the corresponding modulation period and on the correlation length. These facts point to the persistence of the modulated structure emerging in the ground state even if, strictly speaking, the on-site order is completely lost in the thermodynamic limit~\cite{Landau_Lifshitz}. 
In Section~\ref{discussion}, we provide some arguments aiming at explaining, within an analytic approach, the cross-over from the $g=0$ Ising universality class to a continuous-symmetry behavior for $g\geq 0$ and $\alpha \le \alpha_0$
as well as the temperature dependence of some physical observables in comparison with MC simulations. 
In Section~\ref{conclusions}, we provide a summary of the most relevant results and indicate possible directions for further work. Technical aspects of the MC simulations and of the analytical computations are presented in Appendices.

\section{Monte Carlo results}
\label{gsp}
%%%%%%%%%%%%%%%%%%%%%%%%%%%%%%%%%%%%%%%%%%%
\subsection{The model and the ground state}
The Hamiltonian with Ising variables $\sigma_i=\pm 1$ on a 1d lattice reads 
\begin{equation} 
\label{DFIF_Hamiltonian} 
\mathcal{H}=
-J\sum_{j=1}^N \sigma_j  \sigma_{j+1} 
+\frac{g}{2}\sum_{\{ i \ne j \}}\frac{\sigma_i  \sigma_{j}}{|i-j|^{\alpha} }\,,
\end{equation}
where $N$ is the number of spins in the chain and $\{ i\ne j \}$ indicates a sum over all the couples in the chain; 
periodic boundary conditions $\sigma_{i+N}=\sigma_i$ are assumed.  
The ground state~\cite{Lieb} of this Hamiltonian is uniform for $\alpha> \alpha_0$, where $\alpha _0\geq 2$ depends on the ratio $J/g$. For $\alpha \le \alpha_0$, the ground state consists of a regular sequence of groups of $h$ adjacent spins with positive ($\sigma_j=+1$) and negative ($\sigma_j=-1$)  orientation. 
\begin{figure} % FIGURE 1
  \begin{center}               
     \includegraphics[width=0.42\textwidth]{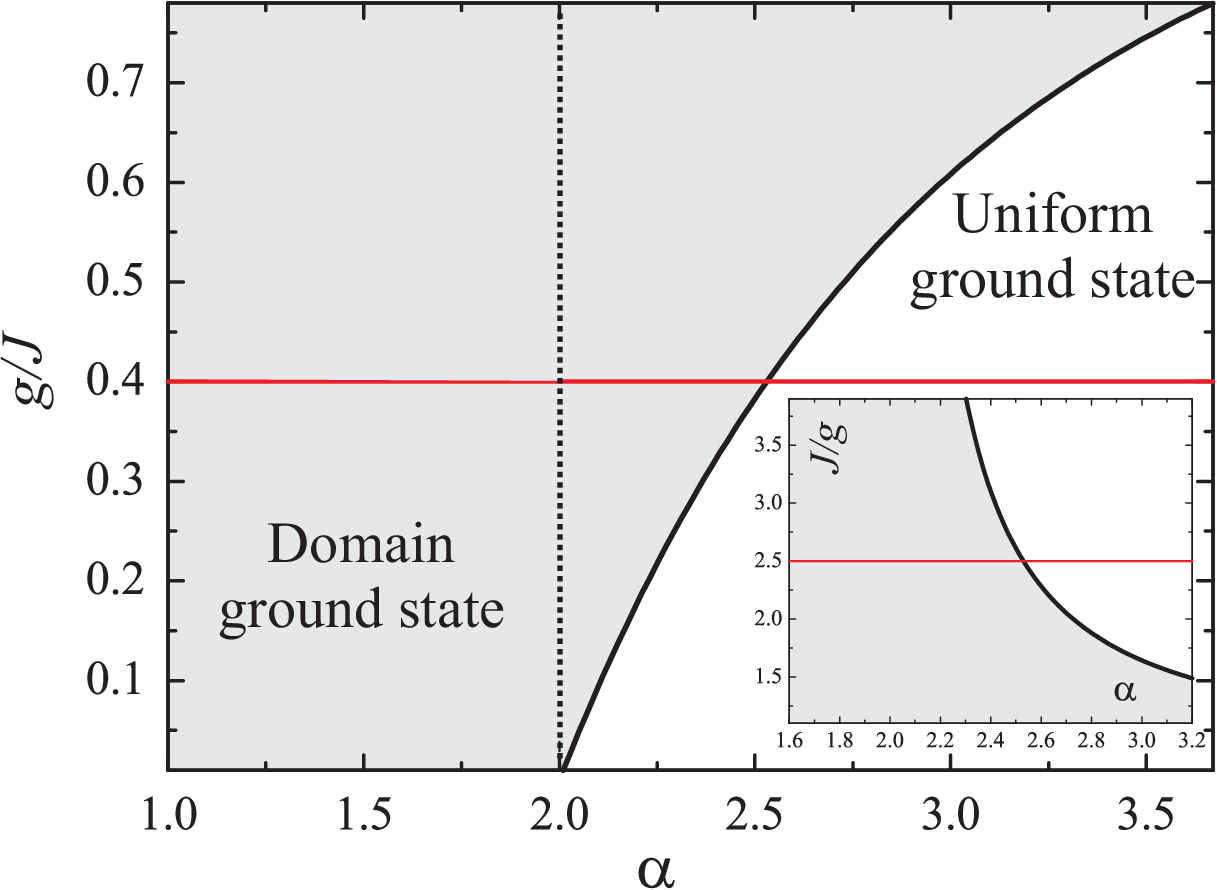}
  \end{center}
  \caption{(Color online) Ground-state phase diagram in the $\left(\alpha\,,g/J\right)$ plane. For $1\le\alpha\le 2$ the 
    ground state always consists of domains (grey region). For $\alpha > 2$, 
    the crossover to a uniform ground state (white region) occurs when 
    $\alpha$ exceeds the threshold value $\alpha_0$ indicated by the solid line. 
    MC calculations have been performed for values of $\alpha$ between 1.6 and 3.2 
    and $g/J=0.4$ (horizontal red line). Inset: zoom of the region where MC simulations have 
    been performed in the $\left(\alpha\,,J/g\right)$ plane, $J/g=2.5$ (horizontal red line).}       
  \label{fig_1}
\end{figure} 
The zero-temperature phase diagram in the $\left(\alpha\,,g/J\right)$ plane  
is schematically reported in Fig.~\ref{fig_1}. The inset zooms into the region of the parameter space 
($\left(\alpha\,,J/g\right)$ plane in this case) in which MC simulations have been performed: $J/g=2.5$ and $\alpha= 1.6 \dots 3.2$.  
The main thermodynamic observable we address is the two-point correlation function at temperature $T$ and fixed $\alpha$:
\begin{equation}
\label{corrfun_MC}
C_\alpha(r)= \langle \langle \sigma_{j+r} \sigma_{j} \rangle_j\rangle_T 
\end{equation}
and its Fourier transform $\mathcal{S}_\alpha(q)$ (commonly named structure factor):
\begin{equation}
\label{strfac_MC}
\mathcal{S}_\alpha(q) = \sum_{r=-\infty}^{+\infty}\langle \langle \sigma_{j+r} \sigma_{j} \rangle_j\rangle_T e^{-iqr}\,.
\end{equation}
As the system cannot be assumed to have translational invariance, an average over the lattice sites $j$ is needed  
($\langle \dots \rangle_j$ in \eqref{corrfun_MC}, \eqref{strfac_MC} and henceforth); 
$\langle \dots \rangle_T$ denotes the thermal average. 
The physical quantities computed with the MC approach actually correspond  to the double average $\langle\dots \rangle=\langle \langle\dots \rangle_j\rangle_T$.

The lowest-energy spin profiles are known to be square waves ${\rm Sq}(k_0 j) $, with a   
modulation period $2h = 2\pi/k_0$~\cite{Lieb}. The total energy can be parameterized with $h$ by inserting the square profile into the Hamiltonian~\eqref{DFIF_Hamiltonian}. The ground state equilibrium value of $h$ -- let us call it $h_{gs}$, corresponding to $k_{gs}$ -- is then determined by minimizing the resulting energy~\eqref{DFIF_gs_energy_app_B} with 
respect to $h$. $h_{gs}$ depends on $\alpha$ and $J/g$: some values are reported in Fig.~\ref{fig_5}. The two-point correlation function for a generic square-wave profile reads (see Appendix~\ref{appendix_B} for details):  
\begin{eqnarray}
\langle \sigma_{j+r}\sigma_{j}\rangle_j &=& \frac{1}{N}\sum_{j=1}^N{\rm Sq}(k_0 (j+r)) {\rm Sq}(k_0 j) \nonumber\\
&=&\frac{1}{2}\sum_{m=0}^{\infty} a^2_m \cos\left(k_m r \right)\doteq {\rm Tr}(k_0 r)\, , \nonumber\\\label{GS_ss_derivation}
\end{eqnarray}
where $k_0=\pi/h$, $k_m =(2m+1)\cdot k_0$, $a_m = 4\left[\pi(2m+1)\right]^{-1}$ 
and ${\rm Tr}(k_0 r)$ is a symmetric triangular wave of period $2h$. According to~\eqref{GS_ss_derivation} evaluated in $h=h_{gs}$, 
the ground-state structure factor only takes non-zero values in the points located at $q=\pm (2m+1)\cdot k_{gs}$ for which 
$\mathcal{S}_\alpha\left((2m+1)\cdot k_{gs}\right) = N \cdot 4\left[\pi(2m+1)\right]^{-2}$. The structure factor of a uniform state takes a finite value at $q=0$ only: $\mathcal{S}_\alpha(0)=N$.  
This case can be regarded as the limit $h\rightarrow\infty$ so that $k_0=0$ and all 
the peaks of $\mathcal{S}_\alpha(k_m)$ collapse into the peak at $q=0$. 

\subsection{Finite temperature}

\begin{figure} % FIGURE 2
  \begin{center}         
     \includegraphics[width=0.45\textwidth]{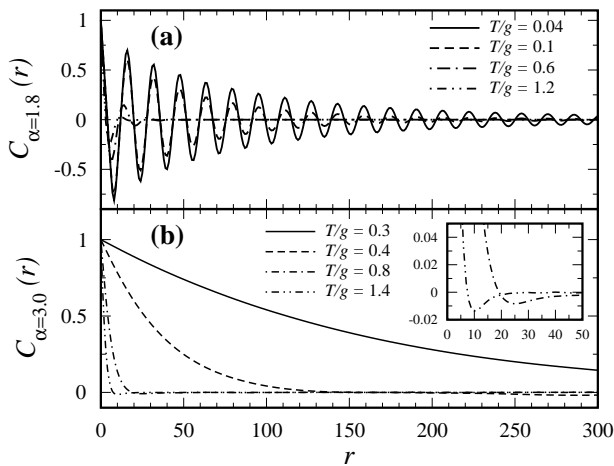}
  \end{center}
  \caption{\textbf{(a)} Correlation function for $\alpha=1.8$ (domain ground state) 
    at different temperatures for $L=1000$ and $J=2.5$. \textbf{(b)} Correlation function for 
    $\alpha=3.0$ (uniform ground state) at different temperatures for $L=1000$ and $J=2.5$.
    Inset: Reminiscence of the competing dipolar interaction (see text).}
  \label{fig_2}
\end{figure}
Fig.~\ref{fig_2} shows the two-point correlation functions \eqref{corrfun_MC} computed by MC simulations for $\alpha=1.8$ (domain ground state) and $\alpha=3.0$ (uniform ground state) at different 
temperatures (see Appendix~\ref{appendix_A} for details about the computational methods).  
In spite of the fact that the single-spin average 
$\langle \langle\sigma_j \rangle_j\rangle_T$ is zero at any finite temperature, 
the correlation function reproduces the essential aspects of the ground state spin configurations. 
For $\alpha=1.8$ (Fig.~\ref{fig_2}a), $C_\alpha(r)$ 
displays an oscillatory decay as a function of $r$, indicating that the loss of on-site magnetization proceeds in such a way that the ground-state segment-order is maintained. 
In the regime in which the  ground state is uniform ($\alpha=3$), instead, 
the correlation function decays smoothly and, in general, monotonically  (Fig.~\ref{fig_2}b).   
A closer look at the highest reported temperatures ($T/g=0.8\,,1.4$) reveals a small interval at short distances 
in which $C_\alpha(r)$ becomes negative (inset in Fig.~\ref{fig_2}b).
This might be taken as an indication that, even starting from a uniform ground state, 
when the temperature is increased the system can spontaneously produce a 
phase with reduced symmetry in which the short-range order occurs with a well-defined modulation. 
We will come back to this 
point at the end of Section~\ref{discussion}. 

\begin{figure}  % FIGURE 3
  \begin{center}     
     \includegraphics[width=0.45\textwidth]{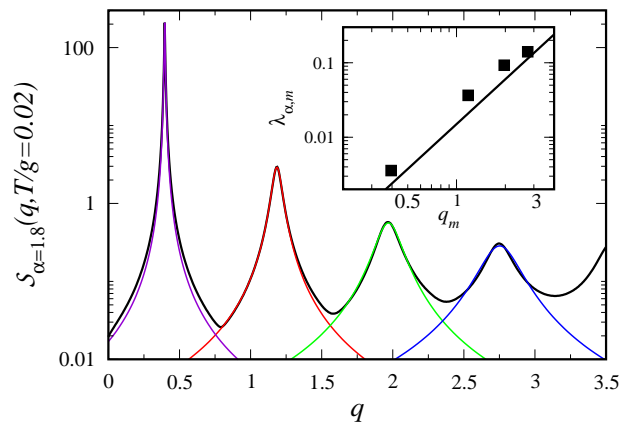}
  \end{center}
  \caption{(Color online) Structure factor obtained by MC 
    simulations at $T/g=0.02$ for $\alpha=1.8$ with Lorentzian 
    fitting for the peaks located at $q=q_m=(2m+1)\cdot q_{\alpha,max}$ (and $J=2.5$). 
    Inset: The HWHM $\lambda_{\alpha,m}$ obtained by Lorentzian fitting of the MC data (full squares) and given 
    by formula~\eqref{lambda_m} (line) is plotted \textit{versus} $q_m$ in a log-log scale, which clearly reveals 
    the power-law behavior.} 
  \label{fig_3}
\end{figure} 
In Fig.~\ref{fig_3}, the structure factor corresponding to 
$\alpha=1.8$ and $T=0.02$ is plotted. The set of discrete 
peaks of the ground state have broadened to Lorentzians centered at $q=(2m+1)\cdot q_{\alpha,max}$. Here, 
$q_{\alpha,max}$ means the position of the highest-peak of the \textit{simulated} 
$\mathcal{S}_{\alpha}(q)$ at finite $T$ and does not, in general, coincide with $k_{gs}$ (the temperature dependence of $q_{\alpha,max}$ will be discussed below). The occurrence of multiple peaks in the finite-temperature structure factor not only indicates that 
the periodic structure of the ground state propagates at finite temperatures but also shows that some  
memory of the detailed square-wave spin profile is retained. As $T$ is increased, peaks with $m>0$ rapidly lose weight and, for $T \gtrsim 0.1$, basically only one peak is detectable. This implies a change of the correlation profile from triangular-wave-like (all harmonics) at low temperatures to cosine-like (single harmonic) at higher temperatures: the same cross-over is predicted to occur for the equilibrium mean-field spin profile within a 2d stripe-domain pattern and observed experimentally in the striped phase of ultra-thin Fe films grown epitaxially on Cu(001)~\cite{Ale}. 
Note that the height of the peaks of $\mathcal{S}_{\alpha}(q)$ in the ground state scales like $(2m+1)^{-2}$, while the ratio between the peaks at $m=1$ and $m=0$ in Fig.~\ref{fig_3} is about one order of magnitude smaller at finite temperatures. In the next Section, we will give a simple explanation for this observation.  The Lorentzian shape of the peaks and the $m$-dependence of their width (inset) -- which are related to the exponential spatial dumping of the correlation shown in Fig.~\ref{fig_2}a -- will also be discussed in Section~\ref{discussion}.

\begin{figure} % FIGURE 4
  \begin{center}               
     \includegraphics[width=0.45\textwidth]{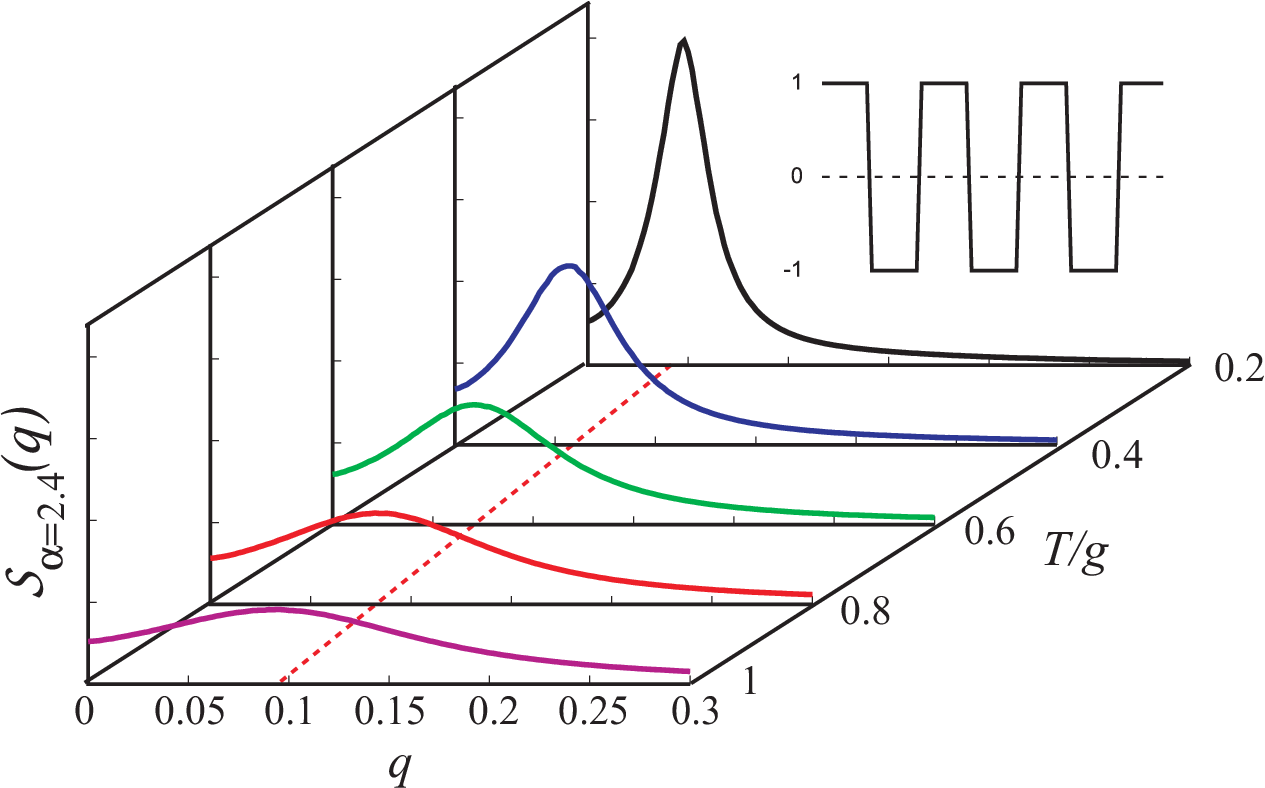}
  \end{center}
  \caption{(Color online) Evolution of $\mathcal{S}_{\alpha}(q)$ as a 
    function of temperature: The red dotted line represents the peak 
    position $q_{\alpha, max}$ at different temperatures 
    (parameters: $J/g=2.5$ and $\alpha=2.4$, giving $h_{gs}=51$). 
    Inset: Schematic view of a square-wave spin profile.}
  \label{fig_4}
\end{figure}
A typical temperature dependence of the structure factor is shown in Fig.~\ref{fig_4}, using a linear scale where only the most prominent peak $m=0$ is evident. Two facts are visible: 
{\it 1.)}  the location of the maximum $q_{\alpha,max}$ varies with temperature and 
{\it 2.)}  the peak broadens considerably when the temperature is increased. 
We will discuss these two features more thoroughly.

\subsubsection{Temperature and $\alpha $-dependence of $q_{\alpha, max}$}
\begin{figure} % FIGURE 5
  \begin{center}               
    \includegraphics[width=0.45\textwidth]{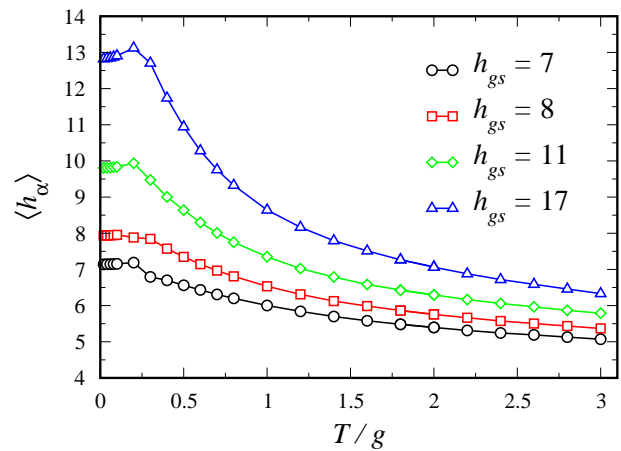}
  \end{center}
  \caption{(Color online) Plot of $\langle h_\alpha\rangle$ \textit{versus} $T/g$ 
    in the domain-ground-state region. The computation parameters are $L=1000$, $J/g=2.5$, 
    and $\alpha$\,=\,1.6 (circles),\,1.8 (squares),\,2 (diamonds),\,2.2  
    (triangles). The statistical errors are smaller than the data symbols.}
  \label{fig_5}
\end{figure}
$\langle h_{\alpha}\rangle \doteq \pi/q_{\alpha, max}$ is plotted as a function of the temperature  
in Fig.~\ref{fig_5} for the set of values $\alpha=1.6,1.8,2,2.2$,  
all in the regime $\alpha<\alpha_0$ for the chosen $J/g$.  
The ground-state value $h_{gs}$, found by minimizing the total 
energy~\eqref{DFIF_gs_energy_app_B}  with respect to $h$, is also indicated. 
When $\alpha $ is increased -- approaching the transition line to the uniform state  -- both ground-state and finite-temperature values also increase. A strongly decaying long-range interaction favors longer periods. 
To be more quantitative, two temperature regions have to be considered: 
\begin{itemize}
\item For $T/J\gtrsim$0.3,  
the period of modulation decreases with temperature, 
in a similar way to what is found for the stripe width in the Mean-Field Approximation (MFA) 
of a similar but 2d model and in line with experimental results~\cite{Ale}. %The low-temperature range 
\item The temperature range $T/J\lesssim$0.3 is more difficult to explain.  
The modulation period saturates at the ground-state value for $\alpha=1.6,1.8$ and remains below the ground-state value  
for $\alpha=2.0,2.2$. We interpret the convergence of 
$\langle h_{\alpha}\rangle \rightarrow h_{gs}$ with $T\rightarrow 0$ 
as a positive indication that our MC calculations capture the essential equilibrium properties of the model, although we 
note that for larger periods, in this temperature range, 
the MC acceptance rate approaches zero (``blocked condition''). 
A further investigation should be required to decide whether 
this is due to a technical limitation or rather to the set-in of 
intrinsic slow dynamics, by analogy with similar 
systems~\cite{Schmalian_Wolynes,Cannas_slow_dyn_1,Cannas_slow_dyn_2,Cannas_slow_dyn_3,Cannas_slow_dyn_4}. \\
\end{itemize} 
In Appendix~\ref{appendix_D} we will introduce an energy functional for finite $T$ which depends
parametrically on the period of modulation $2h$. Within some approximations, 
there we show that the minimum of such a functional is found for smaller and smaller $h$ 
as the temperature is increased, thus reproducing qualitatively the dependence of  $q_{\alpha, max}$ 
on $T$.

\subsubsection{Temperature and $\alpha $-dependence of the correlation length $\xi _\alpha (T)\doteq \lambda_{\alpha,max}^{-1}$, $\lambda_{\alpha,max}$ being the Half Width at Half Maximum (HWHM) of the Lorentzian centered at $q_{\alpha, max}$}
\begin{figure} % FIGURE 6
  \begin{center}               
    \includegraphics[width=0.45\textwidth]{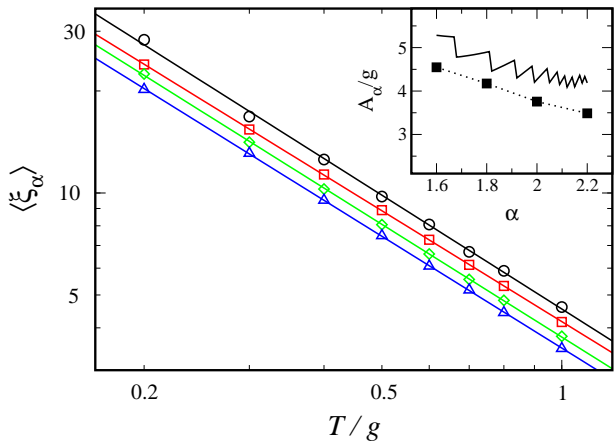} 
  \end{center}
  \caption{(Color online) Plot of $\xi_\alpha$ \textit{versus} $T/g$ 
    in the domain-ground-state region. The computation parameters are $L=$1000, $J/g=2.5$, 
    and $\alpha$\,=\,1.6 (circles),\,1.8 (squares),\,2 (diamonds),\,2.2 
	    (triangles). The continuous lines represent the fit function 
    $y=A_\alpha/x^{B_\alpha}$ (see text). 
    The statistical errors are smaller than the data symbols. 
    Inset: The values of $A_{\alpha}/g$ obtained by fitting the results of the MC 
	    simulations (main frame) are reported as a function of $\alpha$
    (full squares connected by dotted line). The theoretical value of 
    the prefactor of $1/T$ in formula~\eqref{xi_DFF},  
    $(4/\pi^2 ) h^4 \left(\partial ^2 \mathcal{E}_{gs}/\partial h^2\right)$, is indicated by the solid ``zig-zag'' line (see text).}
  \label{fig_6}
\end{figure} 
In Fig.~\ref{fig_6}, $ \xi_\alpha $ is plotted \textit{versus} $T/g$ for 
$\alpha=1.6,1.8,2,2.2$ (all falling in the region $\alpha \le \alpha_0$ for $J/g=2.5$)   
in a log-log scale.  
Dots correspond to MC data while the solid lines represent fits with the function $A_\alpha/T^{B_\alpha}$, with fitting  parameters $A_\alpha$ and $B_\alpha$. 
The best fit yields the same exponent 
$B_\alpha=1.10(5)$ for each $\alpha$, while $A_\alpha$ has a more complicated dependence on $\alpha$, 
see squares in the inset of Fig.~\ref{fig_6} (the zig-zag line will be discussed in the next Section). We conclude that the 
dependence of the correlation length on $T$ is better described by 
$A_\alpha/T^{B_\alpha}$ than the Ising
exponential relation  $e^{\Delta/T}$, which holds for $g=0$. 
A deeper understanding of this difference will be provided in Section~\ref{discussion}. 
For a comparison with the uniform regime (i.e., $\alpha=2.6,2.8,3,3.2$), let us consider  
the  correlation function for $\alpha=3.0$ displayed in Fig.~\ref{fig_2}b.
Note that the plot is limited to low enough temperatures in order to avoid the anomalous range of spatial decay where the correlation function becomes negative (see inset of Fig.~\ref{fig_2}b). 
\begin{figure}  % FIGURE 7
  \begin{center}               
     \includegraphics[width=0.45\textwidth]{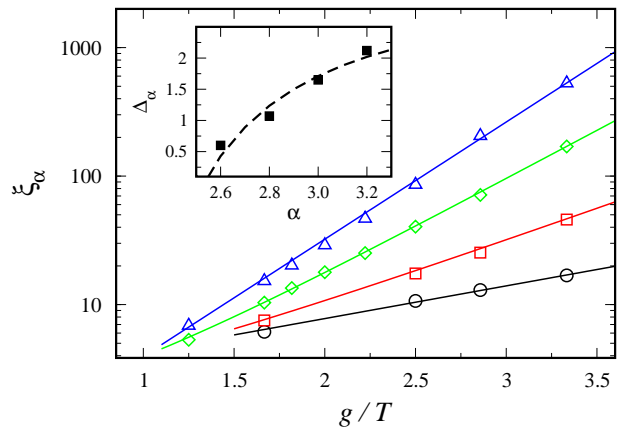}
  \end{center}
  \caption{(Color online) Log-linear plot of $\xi_\alpha$ 
    \textit{versus} $g/T$ in the uniform-ground-state region. The computation parameters are $L=1000$, 
    $J/g=2.5$, and $\alpha$\,=\,2.6 (circles),\,2.8 (squares),
    \,3.0 (diamonds),\,3.2 (triangles). The continuous lines are best fit functions. 
    The statistical errors are smaller than the data symbols. Inset: The dependence 
    of the barrier $\Delta_w=2J-2g \zeta(\alpha-1)$ derived in Sect.~\ref{discussion} on $\alpha$  
    (dashed line) is compared with that of $\Delta_{\alpha}$ obtained by fitting $\xi_\alpha$  computed in  MC 
    simulations (full squares).}
  \label{fig_7}
\end{figure}
Besides, in the temperature range $T =0.3 \dots 0.8$, the $g=0$ behavior is recovered.
In fact, looking at the correlation length $\xi_{\alpha}$
for $\alpha=2.6,2.8,3,3.2$, we find that it is better fitted by 
$\xi_{\alpha} \sim \exp\left(\Delta_{\alpha} /T \right)$, see the log-linear plot of $\xi_{\alpha}$ \textit{versus} $g/T$ in Fig.~\ref{fig_7}.  
Remarkably, the energy barrier $\Delta_{\alpha}$ does depend on $\alpha$,  
see squares in the inset of Fig.~\ref{fig_7} (the dashed line will be discussed at the end of the next Section).

Regarding the possibility to observe  
long-range order at finite temperature,  
our MC results seem to exclude such a hypothesis. 
In fact, in a correct analysis of the structure factor, beyond the usual intensive
term connected with the correlations (Lorenzian-like functions), 
one should also take into account an extensive factor  
associated with the occurrence of long-range order~\cite{chailub}.
This last component has always been considered in the fitting procedure but it has 
never given a significant contribution to 
the simulated structure factors~\eqref{strfac_MC}.  
The occurrence of long-range order at $T=0$ only 
is also supported by the low-temperature divergence of $\xi_\alpha$ 
for both $\alpha\le \alpha_0$ (Fig.~\ref{fig_6}) and $\alpha > \alpha_0$ (Fig.~\ref{fig_7}). 
The whole scenario confirms some recent theoretical works. 
In fact, in Ref.~\onlinecite{kerimov} the absence of long-range 
order  at any temperature for $g > 0$, $J = 0$ and $\alpha >1$ is rigorously proved.
Even if in that specific case a short-range ferromagnetic term was not included, 
%besides it has been subsequently conjectured in Ref.~\onlinecite{Lieb} 
it seems reasonable to extend such a result to the case $J >0$ 
and conclude that long-range order should not occur with the model~\eqref{DFIF_Hamiltonian} 
for $\alpha > 1$~\cite{Lieb}. 
The occurrence of a phase transition has been suggested, instead, for $0<\alpha\le 1$ so that it 
would be particularly interesting to investigate 
the finite-temperature properties of the model~\eqref{DFIF_Hamiltonian} in this regime. 
However, several other issues are related to the  
divergence of the energy per spin for pair-spin interaction decaying like $1/r^\alpha$ 
with $\alpha < d$, $d$ being the dimension of the space in which the spin system is embedded, 
such as energy \textit{non-additivity} and \textit{ensemble inequivalence}~\cite{Ruffo_2,barre05,mukamel} .
For this reason our canonical MC method would not necessarily provide the unique and 
correct results in this context, but further analyses 
involving the comparison of different computational approaches would be, most probably, required.

\section{Discussion}
\label{discussion}
In this Section, we provide an explanation for some of the MC results on the basis of a simple physical model for the excited states of the Hamiltonian~\eqref{DFIF_Hamiltonian}. In particular, we will provide a physical picture for the temperature dependence of the correlation length in the two distinct regimes $\alpha \le\alpha _0$ and $\alpha >\alpha _0$.

\subsection{Case $\alpha \le \alpha _0$}
 We construct excited states of the Hamiltonian~\eqref{DFIF_Hamiltonian} by modifying the square-wave profile to
\begin{equation}
\sigma_j={\rm Sq}\left(k_0 (j+ u_j )\right) =\sum_{m=0}^{\infty}  a_m \sin\left(k_m  (j+ u_j )\right)  
\end{equation}
with $u_j$ being a displacement field. This perturbation corresponds to displacing the position of the wall between adjacent segments, which creates a generally non-periodic spin configuration. The quantity we need to compute is the increment of energy due to the displacement field:
\begin{equation}
\Delta  \mathcal{E}_{h}\doteq \mathcal{E}_{h}[u]-\mathcal{E}_{h}[ u=0]
\end{equation}
with $\mathcal{E}_{h}$ being defined as $\langle \mathcal{H}\rangle_j$.  
$\Delta  \mathcal{E}_{h}$ is computed perturbatively, i.e.,  
in the limit of small $\tilde{u}_q$ ($u_j \doteq \left(1/N\right)\sum_q\tilde{u}_q e^{i q j}$), see Appendix~\ref{appendix_C}. 
For $q<<k_{0}$ and setting $k_0= k_{gs}$, with $k_{gs}$ being $\pi/h_{gs}$, one has 
\begin{equation} 
\label{elastic_DFF}
\Delta  \mathcal{E}_{gs}= \frac{1}{N} \sum_q  
\left[
\frac{1}{2} k_{gs}^2
\frac{\partial ^2 \mathcal{E}_{gs}}{\partial k_{0}^2} 
q^2 |\tilde{u}_q|^2 \right]\,.
\end{equation}   
Eq.~\eqref{elastic_DFF} describes the spectrum of the excited states 
(see also  Refs.~\onlinecite{Kashuba_Pokrovsky,Abanov_Pokrovsky} for a model in 2d) and the coefficient of 
$q^2$ is a stiffness $k_{gs}^2\left(\partial ^2 \mathcal{E}_{gs}/\partial k_0^2\right)$ against fluctuations from the ground-state spin configuration (see Appendix~\ref{appendix_C} for $\partial ^2 \mathcal{E}_{gs}/\partial k_0^2$ definition).   
Note the gapless, quasi-continuum nature of the spectrum of fluctuations, 
in clear contrast to the gapped spectrum of fluctuations in a pure ($g=0$) Ising model.
  
Eq.~\eqref{elastic_DFF} is the central result of this Section, as it allows computing 
the  structure factor $\mathcal{S}_{\alpha}(q)$ and the correlation length $\xi_{\alpha}(T)$. 
The resulting structure factor~\eqref{strfac_MC} consists of  a series of Lorentzian peaks 
centered at $q= \pm k_m$   
\begin{eqnarray}
\mathcal{S}_{\alpha}(q)
&=&\frac{1}{2}\sum_{m=0}^{\infty}\Bigg\{ a^2_m  \nonumber\\
&\times&\left.\left[
\frac{\lambda_{\alpha,m}}{ \left(q-k_m\right)^2 + \lambda_{\alpha,m}^2} 
+ \frac{\lambda_{\alpha,m}}{ \left(q+k_m\right)^2 + \lambda_{\alpha,m}^2}  
\right] \right\} \, ,\nonumber\\
\label{Struct_factor}
\end{eqnarray}
with a HWHM given by 
\begin{equation}
\label{lambda_m}
\lambda_{\alpha,m} 
=(2m+1)^2 \frac{T}{2\frac{\partial ^2 \mathcal{E}_{gs}}{\partial k_0^2} } \,.
\end{equation}
The reader is referred to Appendix~\ref{appendix_C} for the details.  
The same behavior is observed in the MC results plotted in Fig.~\ref{fig_3}. 
Our analysis finds the origin of the multiple peaks of $\mathcal{S}_{\alpha}(q)$ 
in the quasi-continuum spectrum of gapless excitations (see Eq.~\eqref{elastic_DFF})  
appearing in the frustrated model for $\alpha \le \alpha _0$. 
A remarkable feature is the non-trivial scaling of the maxima  with the higher-harmonic index $2m+1$ 
\begin{equation}
\label{eq10}
\mathcal{S}_{\alpha}(q=\pm k_m) = \frac{1}{2}\frac{a_m^2}{\lambda_{\alpha,m}} = \frac{16}{\pi^2}
\frac{\partial ^2 \mathcal{E}_{gs}}{\partial k_0^2} 
\frac{1}{T}\frac{1}{(2m+1)^4} \,,
\end{equation} 
which accounts for the strong reduction detected for the ratio between the 
peak heights for $m=1$ and $m=0$ in the MC results at finite temperatures. 
Note also the square-power dependence of the HWHM $\lambda_{\alpha,m}$ on $2m+1$    
in formula~\eqref{lambda_m}. 
This theoretical prediction (solid line in the inset of Fig.~\ref{fig_3} with log-log scale) 
is in excellent agreement with the behavior of the $\mathcal{S}_{\alpha}(q)$ 
simulated for  $\alpha=1.8$ and $J/g=2.5$  at $T/g=0.02$ (squares in the inset of Fig.~\ref{fig_3}). 
From the assumption $k_{0}=k_{gs}$ (Eq.~\eqref{elastic_DFF}) it follows that, within our analytic model,  
the highest peak of  $\mathcal{S}_{\alpha}(q)$ is expected to occur at $q_{\alpha, max}=k_{gs}$ and 
the correlation length is defined as $\xi_{\alpha} = \lambda_{\alpha,0}^{-1}$ consistently. 
Even if we already know that in MC simulations $q_{\alpha, max}$ 
does not remain constant as $T$ is varied (see Fig.~\ref{fig_5}), this assumption produces 
a $1/T$-dependence of the correlation length 
\begin{equation}
\label{xi_DFF}
\xi_{\alpha}  = \frac{1}{\lambda_{\alpha,0}} = 2  \frac{\partial ^2 \mathcal{E}_{gs}}{\partial k_0^2} \frac{1}{T} \,,
\end{equation}
which is in good agreement ($B_{\alpha}=1.10(5)$) with the 
corresponding quantity computed again with the MC technique (see Fig.~\ref{fig_6}). 
Finally, the analytic model predicts that
$A_\alpha = 2 \left(\partial ^2 \mathcal{E}_{gs}/\partial k_0^2\right)$. 
Computing this expression numerically 
produces the solid curve in the inset of Fig.~\ref{fig_6}.  
The step-like behavior of 
$2 \left(\partial ^2 \mathcal{E}_{gs}/\partial k_0^2\right)$ 
reflects the fact that both 
the optimal domain width $h_{gs}$ and the second derivative of the energy -- computed 
in $h=h_{gs}$ -- are discontinuous functions of $\alpha$~\cite{Lieb} in virtue of the discreteness of the lattice. 
Both the order of magnitude and the scaling with $\alpha $ agree with 
MC calculations (squares in the inset of Fig.~\ref{fig_6}): $A_{\alpha}$ decreases 
as $\alpha$ increases approaching the uniform-ground-state region.
In summary, the agreement between numerical and analytical results indicates that 
the distortion of the ground-state spin profile due to the 
displacement of domain walls represents the main disordering mechanism 
when $\alpha \le \alpha_0$.  \\
To the aim of reproducing the temperature dependence of 
$q_{\alpha,max}$, the expansion for $q \ll k_0$  -- performed in
Appendix~\ref{appendix_C} to get from~\eqref{DFIF_perturb_energy_4_app_B} to~\eqref{DFIF_perturb_energy_5_app_B} -- 
is not expected to be accurate anymore. 
However, in Appendix~\ref{appendix_D} we show that letting 
$h$ be an adjustable parameter at finite $T$ with an appropriate 
(temperature-dependent) stiffness we are able to reproduce qualitatively 
the decrease of the modulation period with increasing temperature observed in MC simulations. 
This, indeed, happens because in the correlation function~\eqref{correlation_DFF_T_app_B}  
higher harmonics are progressively more suppressed as the temperature increases. 
As a result, the competition between the ferromagnetic exchange and 
the antiferromagnetic long-range interaction turns out to be biassed with respect to the zero-temperature case 
and the period of modulation decreases subsequently. 
This close relationship between the suppression of higher-harmonic components 
and the decrease of the characteristic period of modulation has been already highlighted 
experimentally and by mean-field calculations in an equivalent 2d system, suggesting  
that it might be a general property of such models.

The pure Ising Hamiltonian is invariant with respect to any operation that changes the
variable $\sigma _j$ to $- \sigma _j$: it has the discrete symmetry group $\mathbb{Z}_{\,2}$. In the next Subsection, we will discuss this case in connection with 
$\alpha >\alpha _0$. The $1/T$-dependence of the correlation length, obtained by introducing a long-range interaction ($ g \ne 0$), suggests that, in the regime of $\alpha\le\alpha _0$, the frustrated system crosses over to the completely different universality class  of one-dimensional chains hosting a planar spin field with SO(2) continuous symmetry~\cite{Wegner,Fisher}.

\subsection{Case $\alpha>\alpha_0$}
In this regime, the ground state is uniform and the Ising universality class
is restored at low enough temperatures, as shown by the correlation length diverging exponentially as $e^{\Delta/T}$, see Fig.~\ref{fig_7}. Specific to this case is that 
$\Delta = \Delta_{\alpha}$, see Inset Fig.~\ref{fig_7}. We try to explain this result by considering that, 
in the pure Ising model ($g=0$),  
the  barrier $\Delta$ equals the energy 
cost to reverse half of the spins starting from a uniform configuration.   
Were the general arguments 
which associate such an energy with the low-temperature expansion of $\xi$~\cite{Shriffer} 
applicable in the presence of long-range interaction,   
the energy of a single wall would be expected to equal $\Delta_{\alpha}$. 
When half of the spins in the chain are reversed, the exchange energy increases
by $2J$. To compute the variation due to the long-range interaction, 
note that this interaction energy is just given by twice the interaction energy between the two parts of the chains lying on opposite sides with respect to the domain wall 
(as the self-energy in each domain remains the same before and after the flip of half of the spins). This interaction energy is given by 
\begin{equation}
\label{zeta}
\Delta_g = -2g\sum_{j\ge 0}\sum_{i\ge 1}\frac{1}{|j+i|^{\alpha}}=
-2g\sum_{r\ge 1}\frac{r}{r^{\alpha}}=-2g \zeta(\alpha-1)\,,
\end{equation} 
where $\zeta (x)$ is the Riemann zeta function, while $i$ and $j$ are the site indices of spins lying on opposite 
sides of the domain wall.  
The energy to create a wall becomes explicitly dependent on $\alpha $ and amounts to $\Delta_w=2J-2g \zeta(\alpha-1)$. 
In the inset of Fig.~\ref{fig_7}, 
one can appreciate how this estimate actually reproduces 
both the order of magnitude and the dependence on $\alpha$ of the 
energy barrier of the exponentially diverging $\xi_{\alpha}$ obtained from MC simulations.   
To be  rigorous, one should point out that this approach is not completely justified
in this context since, when a long-range interaction is present,  
the creation of a new domain wall is not statistically  
independent of the number and the location of the pre-existing domain walls 
in the chain; such a hypothesis is indeed a basic assumption to put the correlation length in relationship with the cost to create a single wall
in the system~\cite{Shriffer}. Letting $\alpha$ go to infinity effectively reduces the spin-spin interaction to nearest neighbors only so that 
our system becomes equivalent to the usual Ising model, provided that the exchange interaction is replaced by $J-g$.

At $T=0$, the condition $\Delta_w=0$ defines $\alpha_0$. 
In fact, as soon as  $\Delta_w\le 0 $ the uniform configuration has no more the lowest energy 
and the system prefers to split into domains. 
For a given ratio $J/g$, $\alpha_0$ fulfills the condition  $\zeta(\alpha_0-1)=J/g$. 
Using the integral definition of the Riemann zeta function, the previous condition 
can be rewritten as 
\begin{eqnarray}
\frac{J}{g}&=&\zeta(\alpha_0-1)=
\frac{1}{\Gamma(\alpha_0-1)} \int^{\infty}_{0}dx\frac{x^{\alpha_0-2}}{e^x-1}\nonumber\\
\nonumber\\
&=&\frac{1}{\alpha_0-1}\frac{1}{\Gamma(\alpha_0-1)} \int^{\infty}_{0}dx\frac{x^{\alpha_0-1} e^x}{(e^x-1)^2}\nonumber\\
\nonumber\\
&=&\frac{1}{\Gamma(\alpha_0)} \int^{\infty}_{0}dx\frac{x^{\alpha_0-1} e^{-x}}{(1-e^{-2x})^2}\,;\nonumber\\
\label{discrete_vs_Giuliani}
\end{eqnarray} 
this implicit equation for $\alpha _0$ turns out to be 
exact~\cite{Lieb} (the solution being the solid line in Fig.~\ref{fig_1}).
 
For completeness, we recall that at relatively high temperatures a well-defined 
period of modulation seems to emerge in the correlation function  also 
for  $\alpha>\alpha_0$ (see inset of Fig.~\ref{fig_2}b). 
A na\"ive, but essentially correct, interpretation of the temperature dependence of $q_{\alpha,max}$ 
in the regime $\alpha\le\alpha_0$ suggests that thermal fluctuations effectively 
reduce the ratio $J/g$ (the antiferromagnetic long-range interaction is 
fovored in the competition with the ferromagnetic exchange interaction, which  
finally leads to decrease the modulation period with respect to the $T=0$ case).  
In this sense, one may think that even when the uniform pattern has the minimum energy at $T=0$ 
(e.g. for $J/g=2.5$ and $\alpha = 3$  as in Fig.~\ref{fig_2}b), thermal fluctuations induce 
an effective decrease of the ratio $J/g=2.5$ so that a modulated phase eventually has lower free 
energy at high enough temperatures. 
However, this effect can only be evident if such a crossover occurs when there is 
still enough correlation between spins to develop -- at least -- half-period of modulation, 
i.e. roughly for $\xi_{\alpha} >1/q_{\alpha,max}$.  In fact, if the period of the underlying modulated phase 
is much larger than the correlation length $\xi_{\alpha}$, two-point correlations just display a monotonic decay
as a function of the lattice separation. 
A detailed investigation of this phenomenon would be, indeed, intriguing but it is 
beyond the purpose of the present work.

\section{Conclusions}
\label{conclusions}
The Mean-Field Approximation reported e.g. in Ref.~\onlinecite{Ale} provides some straightforward results concerning ferromagnetic Ising system frustrated by a long-range interaction. However, the MFA fails in one important instance: it predicts that the modulated order in the ground state propagates at finite temperatures up to a second-order transition temperature $T_c$, while the Landau-Peierls instability forbids a finite on-site $\langle\sigma_j \rangle_T$ at any finite temperature~\cite{Landau_Lifshitz,Peierls}. 
On the other side, MC simulations are much more accurate than the MFA, but very difficult to perform under experimentally 
realistic conditions. For instance, the large modulation lengths often observed in experiments are practically inaccessible to MC simulations. We concentrated on a model -- Eq.~\eqref{DFIF_Hamiltonian}  -- that is highly simplified but captures some essential characteristics of some physically relevant two-dimensional frustrated systems. Within this model, we have been able to enlarge the modulation length with respect to full two-dimensional MC simulations~\cite{Cannas_Pighin,Rastelli}. 
With this model, we have obtained a set of results that might help to shed light onto some experimental outcomes. In particular: The modulation length appearing in the ground state is found to remain a characteristic length at finite temperatures, where it appears as the length modulating the oscillatory part of the correlation function. Strikingly, it decays with temperature in a way that is similar to the temperature dependence of the stripe-domain width observed in MFA 
and experimentally on Fe/Cu(001) films~\cite{Ale,Oliver1}. In addition, the spatial profile of the correlation function contains the same kind of higher harmonics appearing in the MFA spin profile, with only one fundamental harmonic remaining at sufficiently high temperatures, as specified within the MFA and found experimentally~\cite{Ale}. 
In contrast to the MFA, which predicts a second-order phase transition also in 1d,   
we do not find any trace of a phase transition -- and this is a major deviation from full two-dimensional MC  
simulations~\cite{Cannas_Pighin,Rastelli} or experimental findings. When the spatial decay of the long-range interaction is too short-ranged, the ground state and the finite-temperature state lose the modulated character and become uniform. Correspondingly, the system crosses over from the universality class proper of 1d systems with continuous symmetry~\cite{Wegner,Fisher} to the standard 1d Ising-like universality class~\cite{Ising,Huang}. \\
For future work, a more accurate treatment of the displacement field $u_j$ beyond the   
$q\ll k_0$ approximation (see Appendix~\ref{appendix_C}) is certainly to be considered.

\begin{acknowledgments}
We would like to thank S. Cannas, A. Rettori, P. Politi, D. Stariolo, N. Saratz and M. G. Pini for fruitful discussions. 
The financial support by ETH Zurich and the Swiss National Science Foundation is acknowledged.   
\end{acknowledgments}

\appendix

\section{Monte Carlo Method}
\label{appendix_A}
In this Appendix, we discuss the technical details of the MC method we used 
to study the finite-temperature properties of the Hamiltonian \eqref{DFIF_Hamiltonian}.
A first important issue for the system under investigation 
is the treatment of finite-size effects. In the presence of long-range
interactions, they need to be handled with particular care both numerically and  
analytically~\cite{Ruffo}. 
Some techniques to tackle the problem numerically are given, for instance, in 
Ref.~\onlinecite{libbinder}.  
We perform our simulations on a system containing $L$ spins and treat the  
long-range effects by replicating many identical copies of the ``simulation box''~\cite{Cannas_Lapilli}. 
More explicitly, the interaction between two spins separated 
by $r$ lattice sites reads 
\begin{equation} 
\label{long-range-naked}
G_{\alpha}(r)=\frac{1}{r^{\alpha}} + 
\sum_{n}\frac{1}{|r+nL |^{\alpha}}\,,
\end{equation}
where the index $n$ accounts for the number $N/L$ of replicated boxes. 
Since we have in mind the thermodynamic limit $N \rightarrow \infty$, for numerical evaluation 
of $G_{\alpha}(r)$ we let $n$ go to $\pm \infty$ in order to account for the copies of the system 
lying on both the left- and right-hand sides of the simulated segment,  
containing just $L$ spins. 
The effective coupling  \eqref{long-range-naked} 
can be rewritten, in a way that is more suitable for computational 
purposes: 
\begin{eqnarray} 
G_{\alpha}(r)&=&\frac{1}{r^{\alpha}} + \sum_{n=\pm 1\dots \pm \infty} \frac{1}{|r+nL |^{\alpha}} \nonumber\\
&=&\frac{1}{r^{\alpha}} + \sum_{n=1}^{\infty} \left[ \frac{1}{|r+nL |^{\alpha}}
+ \frac{1}{|r-nL |^{\alpha}}\right]\nonumber\\
&=&\frac{1}{r^{\alpha}} + \frac{1}{L^{\alpha}}
\left\{\sum_{n=1}^{M} \left[ \frac{1}{|n+\frac{r}{L}|^{\alpha}}
+\frac{1}{|n -\frac{r}{L}|^{\alpha}}\right]\right.\nonumber\\
&+&\left.\sum_{n=M+1}^{\infty} \left[ \frac{1}{|n+\frac{r}{L}|^{\alpha}}+\frac{1}{|n -\frac{r}{L}|^{\alpha}}\right]\right\}\nonumber\\
&\simeq& \frac{1}{r^{\alpha}} + \frac{1}{L^{\alpha}}\left\{\sum_{n=1}^{M} \left[ \frac{1}{|n+\frac{r}{L}|^{\alpha}}+\frac{1}{|n -\frac{r}{L}|^{\alpha}}\right]\right.\nonumber\\
&+&\left.2\sum_{n=M+1}^{\infty} \frac{1}{n^{\alpha}}\right\}
=\frac{1}{r^{\alpha}} + \frac{2\zeta(\alpha)}{L^{\alpha}}\nonumber\\ 
&+&\frac{1}{L^{\alpha}}\sum_{n=1}^{M} \left[\frac{1}{|n+\frac{r}{L}|^{\alpha}}+\frac{1}{|n -\frac{r}{L}|^{\alpha}} -\frac{2}{n^{\alpha}}\right] \,;\nonumber\\
\label{long-range}
\end{eqnarray}
in the third passage we have neglected $r/L$ with respect to $M$; the 
error of the whole approximation can be estimated following Ref.~\onlinecite{Cannas_Lapilli}.
This approximation reduces the main computational task to 
evaluating the finite sum over $n$, which is -- however -- rapidly convergent. 
Finally, the working Hamiltonian, restricted to our simulation box, is given by 
\begin{equation} 
\label{DFIF_Hamiltonian_L} 
\mathcal{H}=
-J\sum_{i=1}^L \sigma_i\sigma_{i+1} + \frac{g}{2} \sum_{i=1}^{L} \sum_{j=1}^{L} \sigma_i\sigma_{j} G_{\alpha}(i-j)\,,
\end{equation}
which descends directly from \eqref{DFIF_Hamiltonian}  
with the replica assumption $\sigma_{i \pm nL}=\sigma_i$ ($n=\pm 1\dots \pm \infty$),   
periodic boundary conditions on the simulation box $\sigma_{L+1}=\sigma_{1}$ and setting $r=|i-j|$. Note that 
the indices $i$ and  $j$ now vary in the range $[1,L]$ and are allowed to be equal, $G_{\alpha}(0)$ 
being representative of the interaction between different spins in the original Hamiltonian 
\eqref{DFIF_Hamiltonian}; 
in this particular case ($r=0$), there is no interaction inside the simulation box 
but the $i$-th spin still interacts with its own copies lying in the different 
replicas, $\sigma_{i \pm nL}$, so that 
\begin{equation}
%\label{long-range}
G_{\alpha}(0)=\sum_{n=\pm 1\dots \pm \infty} \frac{1}{|nL |^{\alpha}}=\frac{2\zeta(\alpha)}{L^{\alpha}}\,.
\end{equation}

The MC simulations have been performed using the Simulated Annealing (SA)~\cite{opsa} paradigm. 
The SA is extensively applied in statistical physics with the
intent to study systems where both the ground-state energy 
and the equilibrium at low temperatures are
inaccessible through the basic Metropolis criterion~\cite{libbinder}. 
Certainly, spin glasses~\cite{spinglass}, frustrated magnetic spin structures~\cite{Diep} and models with 
long-range interactions~\cite{Ruffo} are some typical examples of systems 
where the SA and related methods~\cite{Santoro} are largely exploited. 

We also have to remind  that  in literature some cluster methods were employed  in order to reach 
a correct thermodynamic
equilibrium  for a simple model where ferromagnetic long-range interactions are only present \cite{cluster}.  
However, the strong frustration due to the competition between the antiferromagnetic long-range interactions
and the nearest-neighbor ferromagnetic exchange interaction renders the generalization 
of such cluster MC technique to the present case non-trivial. 
For these reason, we have followed  in this work  the main idea of Kirkpatrick \textit{et al.}~\cite{opsa}.
A random initial configuration (which should be considered 
as a paramagnetic state) is picked up. Subsequently, 
the thermodynamic equilibrium at a high enough temperature $T_0$ is established. 
We remember that the MC steps per spin considered here only comprise  
Metropolis moves at the analyzed temperature. 
$T_0$ is usually chosen in order to have a high MC
acceptance ratio per spin. Then the temperature is decreased gradually 
$T\rightarrow T -\Delta T$ ($\Delta T > 0$), and a
fixed number of MC steps per spin $\tau$ is run, starting with the 
last configuration sampled at the previous higher temperature.
So, the main assumption is to 
force a constant and sufficiently slow cooling rate, defined as $r=\Delta T/\tau$.
We have taken $\Delta T =0.1  \dots  0.001$ and $\tau=1 \dots 5\times 10^5$ 
depending on the studied value of $\alpha$. 
The procedure is completed when the ground state is approached. 

We have considered simulation boxes of size $L$=100, 200, 500, 1000, and 2000. 
After discarding the first 1$\times$10$^5$ MC steps, we have collected between 5$\times$10$^5$ and 1$\times$10$^6$ 
measurements of the thermodynamic observables,   
repeating the simulation for each temperature at least three times. The estimation of the statistical errors 
has been achieved by applying the usual blocking technique~\cite{libbinder}.

\section{Correlations in the ground state} 
\label{appendix_B}
In this Appendix, we compute the two-point correlations for a generic
square-wave spin profile, representative of the regime in which 
the ground state consists of domains: $\alpha \le \alpha_0$. 
The lowest-energy configurations, at $T=0$, are known to be given~\cite{Lieb} by 
square-wave spin profiles 
\begin{equation}
\label{square-wave_app_B}
\sigma_j={\rm Sq}(k_0 j) =\sum_{m=0}^{\infty}  a_m \sin\left(k_m j \right)  
\end{equation}
with $k_0=\pi/h$, $k_m = \pi(2m+1)/h$ and $a_m = 4/\left[\pi\left(2m+1\right)\right]$.
With the orthogonality relation $\sum_{j=1}^N e^{-i(k-k')j}=N\delta_{k,k'}$,  
the two-point correlations averaged over the site variables $j$ can be computed: 
\begin{eqnarray}
&&\langle \sigma_{j+r}\sigma_{j}\rangle_j = \frac{1}{N}\sum_{j=1}^N{\rm Sq}(k_0 (j+r)) {\rm Sq}(k_0 j) \nonumber\\
&&= \frac{1}{N}\sum_{j=1}^N\sum_{m,m'=0}^{\infty} a_{m'}  a_m \sin\left(k_m (j+r) \right)   \sin\left(k_{m'} j \right) \nonumber\\ 
&&=\frac{1}{2}\sum_{m=0}^{\infty} a^2_m \cos\left(k_m r \right)\,.\nonumber \\
\label{GS_ss_derivation_app_B}
\end{eqnarray} 
The Fourier coefficients of the series obtained in the final 
passage of equation~\eqref{GS_ss_derivation_app_B} happen to be the same as for the 
symmetric triangular wave of period $2h$ so that in the text we  
use the compact notation $\langle \sigma_{j+r}\sigma_{j}\rangle_j = {\rm Tr}(k_0 r)$.  \\
Eq.~\eqref{GS_ss_derivation_app_B} allows writing the
energy per spin for a general square-wave profile 
\begin{eqnarray} 
\label{energy_generic_square-wave_app_B}
\mathcal{E}_{h}&=&-J \,{\rm Tr}(k_0 )
+\frac{g}{2} \sum_{m=0}^{\infty} a^2_m 
\sum_{r\ge 1}\frac{ \cos\left(k_m r \right) }{r^{\alpha} }\nonumber\\
&=&\sum_{m=0}^{\infty} a^2_m f_{\alpha} (k_m)\,,\nonumber\\ \label{DFIF_gs_energy_app_B} 
\end{eqnarray}
which depends \textit{parametrically} on the half-period of modulation $h$. 
The ground-state energy for a given ratio $J/g$ and $\alpha$ can be 
obtained by minimizing equation~\eqref{DFIF_gs_energy_app_B}  
with respect to $h$ numerically, which consequently defines the equilibrium domain width $h_{gs}$ 
at $T=0$.  
The exchange term in~\eqref{DFIF_gs_energy_app_B} 
straightforwardly gives $-J \,{\rm Tr}(k_0 )=-J\left(1-2/h\right)$, also 
deducible by counting the number of walls present in the domain configuration  
with modulation period $2h$.   
The function  $f_{\alpha}(k_m)=- \left(J/2\right) \cos\left(k_m  \right) + g/2
\sum_{r\ge 1}\left( \cos\left(k_m r \right) /r^{\alpha} \right) $  
introduced above will be used to write the
perturbed energy in a more compact form.

\section{Perturbative treatment of correlations at finite temperatures} 
\label{appendix_C}
In this Appendix, we develop a perturbative elastic model which allows us to
compute the two-point correlations in the regime $\alpha \le \alpha_0$ at finite temperatures. 
Let us consider a displacement field, $u_j$, of the whole square-wave profile~\eqref{square-wave_app_B}:
\begin{equation}
\sigma_j={\rm Sq}\left(k_0 (j+ u_j )\right) =\sum_{m=0}^{\infty}  a_m \sin\left(k_m  (j+ u_j )\right)  \,.
\end{equation}
To compute how the energy~\eqref{GS_ss_derivation_app_B} is 
modified by the presence of this elementary perturbation, we introduce the constants 
\begin{equation}
\begin{cases}
&a=k_m(j+r)\\
&b=k_{m'} j 
\end{cases}
\quad\quad\quad
\begin{cases}
&\gamma=k_m u_{j+r}\\
&\beta=k_{m'} u_j 
\end{cases}
\end{equation}
where the two Greek letters will henceforth be assumed infinitesimal. 
Eq.~\eqref{GS_ss_derivation_app_B} then involves terms like
\begin{eqnarray}
&&\sin(a+\gamma)\sin(b+\beta)=\sin a\sin b \cos\gamma\cos\beta \nonumber\\
&+& \sin a\cos b \cos\gamma\sin\beta +\cos a\sin b \sin\gamma\cos\beta \nonumber\\
&+& \cos a\cos b \sin\gamma\sin\beta \,.\nonumber\\
\label{sin_sin_prod_u_app_B}
\end{eqnarray}
We will further assume that the average over the  
lattice indices $j$, $\langle\dots\rangle_j$, can be performed independently for the 
\textit{rigid} pattern variables (Latin letters)  and for the \textit{fluctuating} 
displacement field $u_j$ 
\footnote{If Eq.~\eqref{sin_sin_prod_u_app_B} were expanded for small 
$u_j \sim \sum_q \tilde{u}_q e^{iqj}$~\eqref{Fourier_displacement_app_B} at this stage, 
the average $\langle \dots \rangle_j$ would produce terms like $\delta(k_m-k_{m'}+q-q')$. However, 
since a further thermal average has to be performed over the variables $u_j$,  
%with $\tilde{u}_q$ obeying a Gaussian distribution, 
this would eventually bring a term $\delta(q-q')$, thus justifying the present factorization of the 
average $\langle\dots\rangle_j$ over the ``Latin'' ($a$ and $b$) and the ``Greek'' ($\gamma$ and $\beta$) 
variables. }: 
\begin{equation}
\begin{split}
\label{j_average_approx_app_B}
\langle \sin a\sin b \cos\gamma\cos\beta \rangle_j =
\langle \sin a\sin b \rangle_j \langle \cos\gamma\cos\beta \rangle_j\,.
\end{split}
\end{equation}
The average  $\langle\dots\rangle_j$ for elementary trigonometric functions 
with arguments $a$ and $b$ gives: 
\begin{equation}
\begin{cases}
&\langle\sin a \sin b\rangle_j = \langle\cos a \cos b\rangle_j =  
\frac{1}{2} \delta_{m,m'} \cos \left(k_m r  \right) \\
&\langle\sin a \cos b\rangle_j = -\langle\cos a \sin b\rangle_j =
\frac{1}{2} \delta_{m,m'} \sin \left(k_m r  \right)  \,,
\end{cases}
\end{equation}
which can be exploited to average~\eqref{sin_sin_prod_u_app_B} 
with respect to $j$: 
\begin{eqnarray}
&&\langle \sin(a+\gamma)\sin(b+\beta)\rangle_j \nonumber\\
&& = \frac{1}{2} \delta_{m,m'} \cos \left(k_m r  \right)
\langle
\cos\gamma\cos\beta + \sin\gamma\sin\beta 
\rangle_j \nonumber\\
&&+\frac{1}{2} \delta_{m,m'} \sin \left(k_m r  \right)
\langle
\cos\gamma\sin\beta - \sin\gamma\cos\beta 
 \rangle_j \nonumber\\
&&=\frac{1}{2} \delta_{m,m'} 
\left\{
\cos \left(k_m r \right)\langle\mathcal{R}e\left[e^{i(\beta-\gamma)}\right] \rangle_j \right.\nonumber\\
&&\left. + \sin \left(k_m r \right)\langle\mathcal{I}m\left[e^{i(\beta-\gamma)}\right] \rangle_j 
\right\}\,;\nonumber\\
\end{eqnarray}
then, recalling that $\beta-\gamma=k_m\left(u_{j}-u_{j+r}\right)$, we get: 
\begin{align}
\langle \sigma_{j+r}\sigma_{j}\rangle_j 
&=\frac{1}{2}\sum_{m=0}^{\infty} 
\left\{ a^2_m 
\left( \cos \left(k_m r\right) \langle\mathcal{R}e\left[e^{i k_m\left(u_{j}-u_{j+r}\right)} \right] \rangle_j \right.\right.\nonumber\\
& \qquad\qquad+ \left.\left.\sin \left(k_m r  \right)\langle\mathcal{I}m\left[e^{i k_m\left(u_{j}-u_{j+r}\right)}\right] \rangle_j  \right)  
 \right\}\nonumber\,.\\\label{ss_correlations_j_app_B}
\end{align}
The introduction of the displacement field brings an increment to 
the energy of a general square-wave profile~\eqref{DFIF_gs_energy_app_B} equal to
\begin{widetext}
\begin{eqnarray} 
\Delta  \mathcal{E}_{h}&=&-J \frac{1}{2} \sum_{m=0}^{\infty} \langle a^2_m 
\left[\sin \left(k_m  \right) \sin\left[k_m(u_j-u_{j+1})\right] \right.
\left.+ \cos \left(k_m \right)
\left(\cos\left[k_m(u_j-u_{j+1})\right]-1\right)
\right] \rangle_j\nonumber\\
&+&\frac{g}{2} \sum_{m=0}^{\infty} \langle a^2_m 
\sum_{r\ge 1}
\left[ \frac{\sin \left(k_m r \right)}{r^{\alpha}} \sin\left[k_m(u_j-u_{j+r})\right] \right.
\left.+ \frac{\cos \left(k_m r \right)}{r^{\alpha}}  \left(\cos\left[k_m(u_j-u_{j+r})\right]-1\right)
\right] \rangle_j \nonumber\\
&\simeq&-J \frac{1}{2} \sum_{m=0}^{\infty} 
\langle a^2_m 
\left[-\sin \left(k_m  \right) k_m(u_{j+1}-u_j) \right.
\left.-\frac{1}{2}  \cos \left(k_m \right)
k_m^2\left(u_{j+1}-u_j\right)^2
\right] 
\rangle_j\nonumber\\
&+&\frac{g}{2} \sum_{m=0}^{\infty} 
\langle a^2_m 
\sum_{r\ge 1}
\left[- \frac{\sin \left(k_m r \right)}{r^{\alpha}} k_m(u_{j+r}-u_j)\right.
\left.- \frac{1}{2} \frac{\cos \left(k_m r \right)}{r^{\alpha}}  
k_m^2\left(u_{j+r}-u_j\right)^2
\right] 
\rangle_j\,,\nonumber\\
\label{DFIF_perturb_energy_app_B} 
\end{eqnarray} 
\end{widetext}
where we have expanded the energy for small displacement differences $u_{j+r}-u_j$.

To proceed in our calculation it is convenient to 
express the displacement field in terms of its Fourier transform $\tilde{u}_q$:
\begin{equation}
\label{Fourier_displacement_app_B}
u_j = \frac{1}{N} \sum_q \tilde{u}_q e^{i q j}
\quad \quad {\rm with} \quad \quad 
\tilde{u}_q = \sum_j u_j e^{-i q j}\,,
\end{equation}
the sum is performed over the Fourier wave numbers $q_m = \pm \left(2\pi m\right)/N $ with $m\in[-N/2,N/2]$, but we drop the index $m$ for simplicity. 
From Eq.~\eqref{Fourier_displacement_app_B} it follows that the averaged square difference is 
\begin{equation}
\label{Fourier_diff_displacement_app_B}
\langle \left( u_{j+r}-u_j\right)^2 \rangle_j = 
\frac{2}{N} \sum_{q} |\tilde{u}_{q}|^2 \left[ 1-\cos(qr)\right]\,,
\end{equation}
while $\langle u_{j+r}-u_j\rangle_j=0$. 
The previous results  and the elementary trigonometric relation 
$\cos(x)\cos(y)= \left(1/2\right)\left[ \cos(x-y)+\cos(x+y)\right]$ 
allow writing~\eqref{DFIF_perturb_energy_app_B}  as  
\begin{widetext}
\begin{eqnarray} 
\Delta  \mathcal{E}_{h}&=&
\frac{1}{N} \sum_{q}  \frac{J}{2}
 \sum_{m=0}^{\infty} \left\{a^2_m 
 k_m^2 \left[
\cos (k_m ) -\frac{1}{2} \cos (k_m -q) 
- \frac{1}{2}   \cos (k_m + q) \right]|\tilde{u}_q|^2
\right\}\nonumber\\
&-&\frac{1}{N} \sum_{q}\frac{g}{2}   \sum_{m=0}^{\infty}
 \left\{ a^2_m  k_m^2  \frac{1}{2}
\sum_{r\ge 1}
\frac{1}{r^{\alpha}}\left[
\cos (k_m r) \right.
\left.-\frac{1}{2} \cos \left[(k_m -q)r\right] -\frac{1}{2}   \cos \left[(k_m + q)r\right]  
\right] |\tilde{u}_q|^2
 \right\}\,.\nonumber\\
\label{DFIF_perturb_energy_3_app_B} 
\end{eqnarray} 
\end{widetext}

Recalling the definition of $f_{\alpha} (k_m)$~\eqref{DFIF_gs_energy_app_B} 
one can  rewrite the perturbed energy~\eqref{DFIF_perturb_energy_3_app_B} as
\begin{eqnarray}
\Delta  \mathcal{E}_{h}&=&
\frac{1}{N} \sum_{q}
\sum_{m=0}^{\infty} \left\{ 
a_m^2 k_m^2 \left[ 
\frac{1}{2} f_{\alpha} (k_m -q) \right.\right.\nonumber\\ 
&+&\left.\left. \frac{1}{2} f_{\alpha}(k_m + q)-f_{\alpha} (k_m )
\right] |\tilde{u}_q|^2
\right\}\,.\nonumber\\
\label{DFIF_perturb_energy_4_app_B} 
\end{eqnarray} 
As far as the large-distance behavior is concerned -- like for the computation of the 
correlation length -- one can expand the energy~\eqref{DFIF_perturb_energy_4_app_B}  for $q \ll k_0$ to get 
\begin{equation} 
\label{DFIF_perturb_energy_5_app_B} 
\Delta  \mathcal{E}_{h}=
\frac{1}{N} \sum_{q} 
\sum_{m=0}^{\infty} \left[
k_m^2 \frac{1}{2} \frac{\partial^2 f_{\alpha} }{\partial^2 k_m} q^2|\tilde{u}_q|^2 
\right]\,,
\end{equation} 
where the derivatives are formally defined by assuming $k_m$ to be a continuum 
variable $k$ and taking the limit 
$\partial^n f_{\alpha} / \partial k_m ^n = \lim _{k\rightarrow k_m} \left(\partial^n f_{\alpha} / \partial k^n\right) $,  
which is, of course, more justified the larger $h$ is. 
The fact that  $\partial k_m/\partial k_0= k_m/k_0$ and   
$\partial^2 k_m /\partial^2 k_0  = 0    $
implies
\begin{eqnarray} 
\frac{\partial ^2 \mathcal{E}_{h}}{\partial k_0^2}&=& 
\sum_{m=0}^{\infty}\left[ 
\frac{\partial^2 k_m }{\partial^2 k_0}  \frac{\partial f_{\alpha} }{\partial k_m } 
+\left(\frac{\partial k_m }{\partial k_0}\right)^2  \frac{\partial^2 f_{\alpha} }{\partial^2 k_m } 
\right]\nonumber\\
&=&\sum_{m=0}^{\infty} \left(\frac{k_m}{k_0}\right)^2  
\frac{\partial^2 f_{\alpha} }{\partial k_m ^2} \nonumber\\
\end{eqnarray} 
so that the perturbed energy~\eqref{DFIF_perturb_energy_5_app_B}  finally reads
\begin{equation} 
\label{elastic_DFF_app_B}
\Delta  \mathcal{E}_{h}= \frac{1}{N} \sum_q  
\left[
\frac{1}{2} k_{0}^2
\frac{\partial ^2 \mathcal{E}_{h}}{\partial k_0^2} 
q^2 |\tilde{u}_q|^2 \right]\,.
\end{equation} 
Eq.~\eqref{elastic_DFF_app_B},  specialized to $h=h_{gs}$ for the ground-state energy,  
essentially matches the result obtained in Ref.~\onlinecite{Kashuba_Pokrovsky} for a 2d system with
an analogous Hamiltonian.   
Within the range of validity of Eq.~\eqref{elastic_DFF_app_B} and with restriction to  $h=h_{gs}$,  
an analytical formula for the structure factor~\eqref{strfac_MC} can be derived. 
The thermal averages $\langle\dots\rangle_T$  of the displacement field $u_j$, 
which appears in the perturbed two-point correlations~\eqref{ss_correlations_j_app_B},  
have to be performed first.   
Those thermal averages can easily be evaluated since the 
Hamiltonian~\eqref{elastic_DFF_app_B} is quadratic for small perturbations of the ground state ($h=h_{gs}$). 
The well-known theorem for Gaussian distributed physical quantities~\cite{Landau_Lifshitz} 
readily gives:   
\begin{eqnarray}
\langle \langle \sigma_{j+r}\sigma_{j}\rangle_j \rangle_T 
&=&\frac{1}{2}\sum_{m=0}^{\infty} \left\{ 
a^2_m \cos \left(k_m r \right) \right.\nonumber\\
&\times&\left. \exp\left[-\frac{1}{2}k_m^2 \langle\langle\left( u_{j}-u_{j+r}\right)^2\rangle_j\rangle_T \right] 
\right\}\,.\nonumber\\
\label{correlation_DFF_T_1_app_B}
\end{eqnarray}
On top of the site average~\eqref{Fourier_diff_displacement_app_B} one has to perform the 
thermal average  
\begin{equation}
\label{Thermal_average}
\langle\langle \left( u_{j+r}-u_j\right)^2 \rangle_j \rangle_T 
= \frac{2}{N} \sum_{q} \langle\tilde{u}_{q}^2\rangle_T \left[ 1-\cos(qr)\right]\,;
\end{equation}
in particular $\langle\tilde{u}_q^2\rangle_T $ can be computed applying the equipartition
theorem to~\eqref{elastic_DFF_app_B}:  
$\left(k_{gs}^2/2\right) \left(\partial ^2 \mathcal{E}_{gs}/\partial k_0^2\right) 
\langle \tilde{u}_q^2 \rangle_T q^2 =  H.~T/2 $     
so that 
\begin{equation}
\label{Fourier_uu_corr_T_1_app_B}
\langle\langle \left( u_{j+r}-u_j\right)^2 \rangle_j \rangle_T =
\frac{T}{k_{gs}^2 \frac{\partial ^2 \mathcal{E}_{gs}}{\partial k_0^2} }
\frac{2}{N} \sum_q \frac{1-\cos(qr)}{q^2}\,.
\end{equation}
By writing the wave numbers explicitly, the sum in the previous formula can be 
evaluated analytically in the thermodynamic limit ($N\rightarrow \infty$): 
\begin{equation}
\begin{split}
&\frac{2}{N} \sum_q \frac{1-\cos(qr)}{q^2}
=\frac{2}{N} \sum_{m=-N/2}^{N/2} \frac{1-\cos\left(\frac{2\pi r}{N}m\right)}{\left(\frac{2\pi r}{N}m\right)^2} \\
&=\frac{N}{2\pi^2}\left\{\frac{1}{2} \left(\frac{2\pi r}{N}\right)^2
+2 \sum_{m=0}^{N/2} \frac{1}{m^2} -2 \sum_{m=0}^{N/2} \frac{\cos\left(\frac{2\pi r}{N}m\right)}{m^2} \right\}\\
&\simeq\frac{N}{2\pi^2}\left\{\frac{1}{2} \left(\frac{2\pi r}{N}\right)^2
+2 \sum_{m=0}^{\infty} \frac{1}{m^2} -2 \sum_{m=0}^{\infty} \frac{\cos\left(\frac{2\pi r}{N}m\right)}{m^2} \right\}\\
&=\frac{N}{2\pi^2}\left\{\frac{1}{2} \left(\frac{2\pi r}{N}\right)^2 
+2\left[\frac{\pi^2}{6} - \frac{1}{4}\left(\frac{2\pi r}{N}\right)^2 \right.\right.\\
&\qquad\left.\left.+\frac{\pi}{2} \left(\frac{2\pi r}{N}\right)-\frac{\pi^2}{6} \right]\right\} 
=\frac{N}{2\pi^2} \frac{2\pi^2 r}{N} = r\,.
\end{split} 
\end{equation}
The thermal average \eqref{Thermal_average} is finally
obtained
\begin{equation}
\label{Fourier_uu_corr_T_app_B}
\langle\langle \left( u_{j+r}-u_j\right)^2 \rangle_j \rangle_T %{\underset{ r\gg 1}{\simeq}} 
=\frac{ T}{ k_{gs}^2 \frac{\partial ^2 \mathcal{E}_{gs}}{\partial k_0^2} }  r\,.
\end{equation}
Combining~\eqref{Fourier_uu_corr_T_app_B} and~\eqref{correlation_DFF_T_1_app_B}, 
the sought-for quantity reads  
\begin{equation}
\label{correlation_DFF_T_app_B} 
\langle \langle \sigma_{j+r}\sigma_{j}\rangle_j \rangle_T 
=\frac{1}{2}\sum_{m=0}^{\infty} \left[ 
a^2_m \cos \left(k_m r \right) e^{- \lambda_{\alpha,m}  r } \right]
\end{equation} 
with 
\begin{equation}
\label{lambda_m_app_B}
\lambda_{\alpha,m} = \frac{1}{2} k_m^2   \frac{T}{k_{gs}^2 \frac{\partial ^2 \mathcal{E}_{gs}}{\partial k_0^2} } 
=(2m+1)^2 \frac{T}{2\frac{\partial ^2 \mathcal{E}_{gs}}{\partial k_0^2} } \,.
\end{equation}
The structure factor~\eqref{strfac_MC} is thus expected to have a series of Lorentzian peaks 
at $q=\pm (2m+1)\cdot k_{gs}$,  $\lambda_{\alpha,m} $ being the corresponding HWHM. More explicitly 
\begin{eqnarray}
\mathcal{S}_{\alpha}(q)
&=&\frac{1}{2}\sum_{m=0}^{\infty} \Bigg\{ a^2_m \nonumber\\
&\times&\left.\left[
\frac{\lambda_{\alpha,m}}{ \left(q-k_m\right)^2 + \lambda_{\alpha,m}^2} + 
\frac{\lambda_{\alpha,m}}{ \left(q+k_m\right)^2 + \lambda_{\alpha,m}^2} 
\right]\right\} \,.\nonumber\\
\label{Struct_factor_app_B}
\end{eqnarray}

\section{Optimal period of modulation at finite temperatures} 
\label{appendix_D}
In this Appendix we provide a qualitative explanation for the dependence of  $q_{\alpha, max}$ 
on the temperature. 
First, we show that the decrease of the modulation period with increasing temperature 
is not reproduced just letting $k_0$ be an adjustable parameter at any temperature. 
In fact, formula \eqref{correlation_DFF_T_app_B} can be used 
to compute the two-point correlations associated with any square-wave profile, provided 
that the appropriate stiffness against deviations from the given 
period $h\ne h_{gs}$ is accordingly employed: $k_{0}^2\left(\partial ^2 \mathcal{E}_{h}/\partial k_0^2\right)$
(see~\eqref{energy_generic_square-wave_app_B} for the definition of $ \mathcal{E}_{h}$).
In this way, one can account for the effect of thermal fluctuations on a square-wave profile 
of an arbitrary half-period $h$ and construct the functional  
\begin{equation} 
\label{H_functional_K_constant} 
\begin{split}
\langle \mathcal{H}_h \rangle&=
-NJ \langle \langle \sigma_{j+1}\sigma_{j}\rangle_j \rangle_{T} 
+N\frac{g}{2} \sum_{r\ge 1} \frac{ \langle \langle \sigma_{j+r}\sigma_{j}\rangle_j \rangle_{T} }{r^\alpha}
\end{split}
\end{equation}
where 
\begin{equation}
\langle \langle \sigma_{j+r}\sigma_{j}\rangle_j \rangle_{T} 
=\frac{1}{2}\sum_{m=0}^{\infty} \left[ 
a^2_m \cos \left(k_m r \right) e^{- \lambda_{\alpha,m}  r } \right]
\end{equation} 
with  $k_m = (2m+1) \pi/h$ ($h\ne h_{gs}$ are here allowed) and 
$\lambda_{\alpha,m} = k_m^2 T/\left[2 k_{0}^2\left(\partial ^2 \mathcal{E}_{h}/\partial k_0^2\right)\right]$. 
The functional $\langle \mathcal{H}_h \rangle$ (Eq.~\eqref{H_functional_K_constant}) can then 
be minimized with respect to $h$ to obtain an 
effective equilibrium period of modulation at finite temperatures. 
\begin{figure}  % FIGURE 8
  \begin{center}               
     \includegraphics[width=0.45\textwidth]{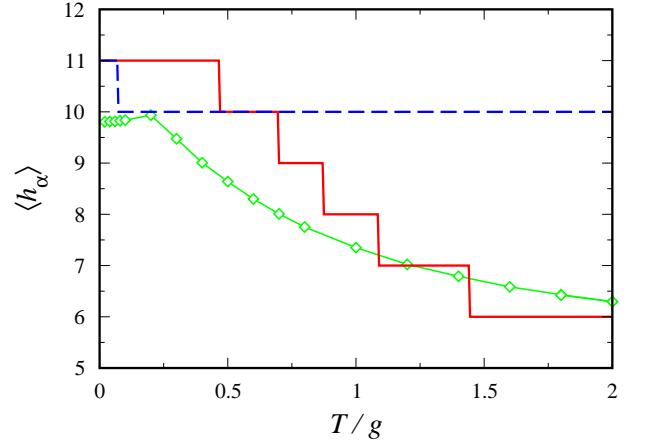}
  \end{center}
 \caption{(Color online) Plot of $\langle h_\alpha\rangle$ \textit{versus} $T/g$ 
    in the domain-ground-state region with $J/g=2.5$ and $\alpha$\,=\,2:
    MC simulations (diamonds), elastic model with constant (dashed line)  
    and temperature dependent (solid line) stiffness,
    $k_{0}^2\left(\partial ^2 \mathcal{E}_{h}/\partial k_0^2\right)$ and 
    $k_{0}^2\left[\partial^2\langle\tilde{\mathcal{H}}_h\rangle_{T}/\partial k_0^2\right]_{\cos}$, 
    respectively (see text). }
  \label{fig_8}
\end{figure}
This procedure produces the dashed line in Fig.~\ref{fig_8}:  
For $J/g=2.5$ and $\alpha=2$, the optimal half-period of modulation corresponds the ground-state value, $h_{gs}=11$,  
for $T< 0.07$, while the functional~\eqref{H_functional_K_constant} has a minimum in $h=10$ 
for higher temperatures. All this indicates that 
the constant decrease of the modulation period observed in the MC simulations  
is not reproduced just by including thermal fluctuations through a displacement field   
into the different square-wave profiles 
and by further minimizing the functional~\eqref{H_functional_K_constant}
with respect to $h$. 
Such a failure might be due to the assumption that the stiffness $k_{0}^2\left(\partial ^2 \mathcal{E}_{h}/\partial k_0^2\right)$ 
remains the same at any temperature. 
In order to circumvent this limitation, we propose a heuristic extension of our elastic model. 
Let us first compute the two-point correlations at an infinitesimal temperature $\delta T$:
\begin{equation}
\label{correlation_dT} 
\langle \langle \sigma_{j+r}\sigma_{j}\rangle_j \rangle_{\delta T} 
=\frac{1}{2}\sum_{m=0}^{\infty} \left[ 
a^2_m \cos \left(k_m r \right) e^{- \delta \lambda_{\alpha,m}  r } \right]
\end{equation} 
with  $k_m = (2m+1) \pi/h$ and
$\delta\lambda_{\alpha,m} = k_m^2 \delta T/\left[2 k_{0}^2\left(\partial ^2 \mathcal{E}_{h}/\partial k_0^2\right)\right]$. 
The correlations \eqref{correlation_dT} can be thought of as resulting from a rigid 
spin profile 
\begin{equation}
\label{noisy_profile}
\sigma_j =\sum_{m=0}^{\infty}  a_m \sin\left(q_m  j \right)  
\end{equation}
in which the wave numbers $q_m=(2m+1)q_0$ are statistically distributed. 
In particular, if a Lorentzian distribution 
\begin{equation}
\label{Lorentzian_q_m_app_B}
P(q_m)= \frac{ \delta\lambda_{\alpha,m} }{\pi} 
\frac{1}{\delta\lambda^2_{\alpha,m}  +\left(q_m - k_m\right)^2}  
\end{equation}
is assumed, the corresponding averages -- performed after the site average $\langle\dots\rangle_j$ --  
mimic the effect of thermal fluctuations such that 
Eq.~\eqref{correlation_dT} can then be rewritten as 
\begin{align}
\langle \langle \sigma_{j+r}\sigma_{j}\rangle_j \rangle_{\delta T} 
&=\langle \langle \sigma_{j+r}\sigma_{j}\rangle_j \rangle_{q_m} \nonumber\\
&=\frac{1}{2}\sum_{m=0}^{\infty} \left[ 
a^2_m \int_{-\infty}^{+\infty} dq_m P(q_m)\cos \left(q_m r \right) \right]\,.
\label{correlation_q_m} 
\end{align}  
The corresponding energy functional reads
\begin{align}
\langle \tilde{\mathcal{H}}_h \rangle_{\delta T}&=\langle \tilde{\mathcal{H}}_h \rangle_{q_m}=
 \int_{-\infty}^{+\infty}dq_m P(q_m) \nonumber\\
&\qquad\times \left[
-NJ \langle \sigma_{j+1}\sigma_{j}\rangle_j 
+N\frac{g}{2} \sum_{r\ge 1} \frac{  \langle \sigma_{j+r}\sigma_{j}\rangle_j }{r^\alpha}
\right]  \nonumber \\
&=\sum_{m=0}^{\infty}  \left[ 
a^2_m \int_{-\infty}^{+\infty} dq_m P(q_m) f_{\alpha}(q_m)  \right] \label{H_functional_q_m} 
\end{align} 
To the aim of computing the correlation function at an infinitesimally 
higher temperature, the spin profile~\eqref{noisy_profile} can be 
further perturbed with a displacement field, which brings an increment 
to the energy functional~\eqref{H_functional_q_m} equal to 
\begin{align}
&\langle \Delta \tilde{\mathcal{H}}_h \rangle_{q_m}
= \frac{1}{N} \sum_q \sum_{m=0}^{\infty} \left\{ 
a^2_m \int_{-\infty}^{+\infty}dq_m P(q_m) \right.\nonumber\\
 &\left.\times q^2_m 
\left[ \frac{1}{2} f_{\alpha} (q_m -q) + \frac{1}{2} f_{\alpha}(q_m + q)-f_{\alpha} (q_m )
\right]  |\tilde{u}_q|^2   \right\}\,.\label{Delta_H_functional_u_q_m} 
\end{align}
By analogy with what done in the previous Section, we perform an expansion 
for $q\ll k_0$ 
(since $q_0$s follow a Lorentzian distribution with maximum in $k_0$, $q \ll q_0$ as well): 
\begin{align} 
\langle \Delta \tilde{\mathcal{H}}_h \rangle_{q_m}&= \frac{1}{N} 
\sum_q \sum_{m=0}^{\infty} \left\{ 
a^2_m \int_{-\infty}^{+\infty} dq_m P(q_m) \right. \nonumber\\ 
&\qquad\qquad\qquad\qquad\qquad\times\left.q^2_m\frac{1}{2} \frac{\partial^2 f_{\alpha} }{\partial q_m^2}  
q^2 |\tilde{u}_q|^2 \right\}  \,.
\end{align}
The fact that $q_m=(2m+1)q_0$ implies $\partial q_m/\partial q_0= q_m/q_0$,   
$\partial^2 q_m /\partial^2 q_0  = 0$ and consequently  
\begin{equation} 
\label{Delta_H_functional_u_small_q_q_m} 
\langle \Delta \tilde{\mathcal{H}}_h \rangle_{q_m}= \frac{1}{N} 
\sum_q \frac{1}{2} \langle q^2_0  \frac{\partial^2\tilde{\mathcal{H}}_h}{\partial q_0^2}  \rangle_{q_m}
q^2 |\tilde{u}_q|^2 \,.
\end{equation}
In the present case, the effective stiffness 
$\langle q^2_0  \left(\partial^2\tilde{\mathcal{H}}_h/\partial q_0^2\right)  \rangle_{q_m}$ 
has a more complicated dependence on $q_0$ with respect to 
Eq.~\eqref{elastic_DFF_app_B}. 
However, we can simplify its computation significantly with the approximation
\begin{equation} 
\label{approx_stiffness_app_B}  
\langle q^2_0  \frac{\partial^2\tilde{\mathcal{H}}_h}{\partial q_0^2}  \rangle_{q_m}\simeq 
 k_0^2  \frac{\partial^2\langle\tilde{\mathcal{H}}_h \rangle_{q_m} }{\partial k_0^2} \big|_{\cos} 
=  k_0^2  \frac{\partial^2\langle\tilde{\mathcal{H}}_h \rangle_{\delta T} }{\partial k_0^2} \big|_{\cos}
\,,
\end{equation}
$\frac{\partial ^2 \langle\tilde{\mathcal{H}}_h\rangle_{\delta T}}{\partial k_0^2}\big|_{\cos}$
meaning that the derivative with respect to $ k_0$ 
involves only the fluctuating functions, $\cos \left(k_m r \right)$. 
The correlation function at  the new temperature ($T= \delta T + \delta T$) is given by 
\begin{align} 
&\langle \langle \sigma_{j+r}\sigma_{j}\rangle_j \rangle_{T} 
=\frac{1}{2}\sum_{m=0}^{\infty}\left[ 
a^2_m \int_{-\infty}^{+\infty} dq_mP(q_m)\right.\nonumber\\
&\qquad\qquad\qquad\times\left.\cos \left(q_m r \right)  
{\rm exp}\left(- \frac{q_m^2}{2}\frac{ \delta T}{ k_0^2  
\frac{\partial^2\langle\tilde{\mathcal{H}}_h \rangle_{\delta T} }{\partial k_0^2} \big|_{\cos} } r \right)  
\right]\nonumber\\
&\simeq \frac{1}{2}\sum_{m=0}^{\infty}\left[ 
a^2_m \int_{-\infty}^{+\infty} dq_mP(q_m)\cos \left(q_m r \right)  \right.\nonumber\\ 
&\qquad\qquad\qquad\qquad\qquad\times\left. {\rm exp}\left(- \frac{k_m^2}{2}\frac{ \delta T}{ k_0^2  
\frac{\partial^2\langle\tilde{\mathcal{H}}_h \rangle_{\delta T} }{\partial k_0^2} \big|_{\cos} } r \right)  
\right]\nonumber\\
\label{correlation_q_m_2}
\end{align}  
where in the last passage we have substituted $q_m^2$ inside the exponential 
with its maximum  $k_m^2$. Such an approximation allows  
writing the energy functional at the new temperature again in the form~\eqref{H_functional_q_m}, provided that  
the HWHM of the Lorentzian distribution $P(q_m)$ is changed into
\begin{equation}
\delta\lambda_m= \frac{k_m^2}{2}\left[ 
\frac{\delta T}{ k_0^2 \frac{\partial^2 \mathcal{E}_h }{\partial k_0^2}} +
\frac{\delta T}{ k_0^2 \frac{\partial^2\langle\tilde{\mathcal{H}}_h \rangle_{\delta T} }{\partial k_0^2} \big|_{\cos} } 
\right]r\,.
\end{equation}
The whole process can then be iterated 
to obtain correlations at any temperature 
\begin{equation}
\langle \langle \sigma_{j+r}\sigma_{j}\rangle_j \rangle_{T} 
=\frac{1}{2}\sum_{m=0}^{\infty} \left[ 
a^2_m \cos \left(k_m r \right) e^{- \lambda_{\alpha,m}(T)  r } \right]\,,
\end{equation} 
and the corresponding energy functional 
\begin{equation} 
\label{H_functional_q_m_2} 
\langle \tilde{\mathcal{H}}_h \rangle_{T} =\langle \tilde{\mathcal{H}}_h \rangle_{q_m}
=\sum_{m=0}^{\infty}  \left[ 
a^2_m \int_{-\infty}^{+\infty} dq_m P(q_m) f_{\alpha}(q_m) \right]
\end{equation}
with HWHM of $P(q_m)$ (letting $\delta T \rightarrow dT$) equal to 
\begin{equation}
\lambda_{\alpha,m}(T)  =k_m^2\int_0^T d\lambda_{\alpha,m} =
\frac{1}{2}\frac{k_m^2}{k_{0}^2} \int_0^T 
\frac{dT}{\frac{\partial^2\langle\tilde{\mathcal{H}}_h \rangle_{T} }{\partial k_0^2} \big|_{\cos} }\,.
\end{equation} 
(Remember that the derivative with respect to $ k_0$ 
involves only the fluctuating functions, $\cos \left(k_m r \right)$, and not the dumping terms). 
By minimizing numerically the functional~\eqref{H_functional_q_m_2} with respect to 
$h$ we obtain the step-like curve in Fig.~\ref{fig_8}. 
In this case a decrease with increasing temperature is indeed observed 
throughout the investigated range. Such a qualitative agreement with 
MC results suggests that the change in the modulation period 
and in the effective stiffness, 
$k_{0}^2\left[\partial^2\langle\tilde{\mathcal{H}}_h\rangle_{T}/\partial k_0^2\right]_{\cos}$, 
should be closely related. 
It is worth remarking that a better agreement is, probably, not to be expected given 
the expansion for $q \ll k_0$ that we performed to pass from 
\eqref{DFIF_perturb_energy_4_app_B}  to \eqref{DFIF_perturb_energy_5_app_B} 
and the further approximations in Eq.~\eqref{approx_stiffness_app_B} and Eq.~\eqref{correlation_q_m_2}.

\end{document}